\documentclass{ws-procs975x65}

\begin{document}
\title{\uppercase{NUMERICAL SIMULATIONS OF A TWO COMPONENT ADVECTIVE FLOW FOR THE STUDY OF THE
SPECTRAL AND TIMING PROPERTIES OF BLACK HOLES}}
\author{KINSUK GIRI$^*$ and SANDIP K. CHAKRABARTI}
\address{S.N.Bose National Centre For Basic Sciences,\\
Block - JD, Sec - III, Salt Lake, Kolkata - 700098, India\\
$^*$E-mail: kinsuk@bose.res.in\\}

\begin{abstract}
Two component advective flows are the most physical accretion disks which 
arise from theoretical consideration. Since viscosity is the determining factor, we investigate the effects 
of viscous stresses on accretion flows around a non­rotating black hole.
As a consequence of angular momentum transfer by
viscosity in an accretion flow, the angular momentum distribution is modified. We include cooling effects and found that a
Keplerian disk is produced on the equatorial plane and the flow above and
below remains sub­ Keplerian. This gives a complete picture to date, of the
formation of a Two component advective flow around a black hole.
\end{abstract}


\bodymatter

\section{Introduction}
Observation evidence of non-thermal photons in the spectrum (Sunyaev \& Truemper,
1979) prompted the model developers to think that a hot electron cloud (the so-called
Compton cloud) along with the standard disk could resolve the issue (Sunyaev \&
Titarchuk, 1980). Numerous suggestions and cartoon diagrams of the illusive
The presence of Compton cloud are shown in the literature (e.g., Zdziarski, 1988 ; Haardt et al.,
1994 ; Chakrabarti \& Titurchuk, 1995 ; Hereafter CT95).
CT95, based on the solutions of viscous and inviscid transonic flows
around black holes (Chakrabarti, 1989; Chakrabarti 1990) proposed that, in general, 
the accretion disk should really have two components: a Keplerian accretion on the equatorial plane and a
sub-Keplerian halo which surrounds the Keplerian disk, and the puffed up inner part
of the flow (CENBOL) which is nothing but the Compton cloud.
There was as yet no work in the literature to show that the Two Component Advective Flow (TCAF) solution is stable.
The cause for concern was obvious: a Keplerian disk is necessarily sub-sonic, while
the sub-Keplerian flow is supersonic, and becomes sub-sonic only at the shock wave.
Thus the questions remained unanswered is: Under what circumstances TCAF actually
forms?  In this paper, through numerical simulations of viscous accretion flow with a power-law cooling effects, we show that
when the injected sub-Keplerian flow angular momentum is high enough and/or the
viscosity is high enough, TCAF would be formed, otherwise the sub-Keplerian flow would remain sub-Keplerian.

\section{Computational Procedure}
All the governing equations for viscous accretion flow have been given in details in
both Giri et al. (2010) and Giri \& Chakrabarti (2012, hereafter GC12) and we will not repeat this once again.
Earlier we have chosen a constant viscosity parameter $\alpha$ in the entire flow as is in vogue in the subject.
In the present work, we have chosen a more realistic $\alpha$ parameter, so that the viscosity is high on the
equatorial plane and low, away from from it. 
The setup of our simulation has been described in GC12. Instead of only viscous flows, 
In the present work, instead of using the same $\alpha$ for the whole $r-z$ plane, we choose a smooth distribution as,
$$
\alpha = {\alpha}_{max} - [{\alpha}_{max}{({z \over {r_{max}}})^{\delta}}],  \eqno(1)
$$
where, $r_{max} = 200, 0 \leq z \leq 200$ and $\delta > 0$. In our cases, we have chosen
$\delta = 1.5$. Clearly, when $z = 0$, i.e. at equatorial plane, $\alpha = {\alpha}_{max} = 0.012$
while, $\alpha = 0$ for $z = z_{max} = r_{max}$. 
We carry out the simulations for several hundreds of dynamical time-scales to achieve a steady state.

\section{Results}
\begin{figure}
\begin{center}
\psfig{file=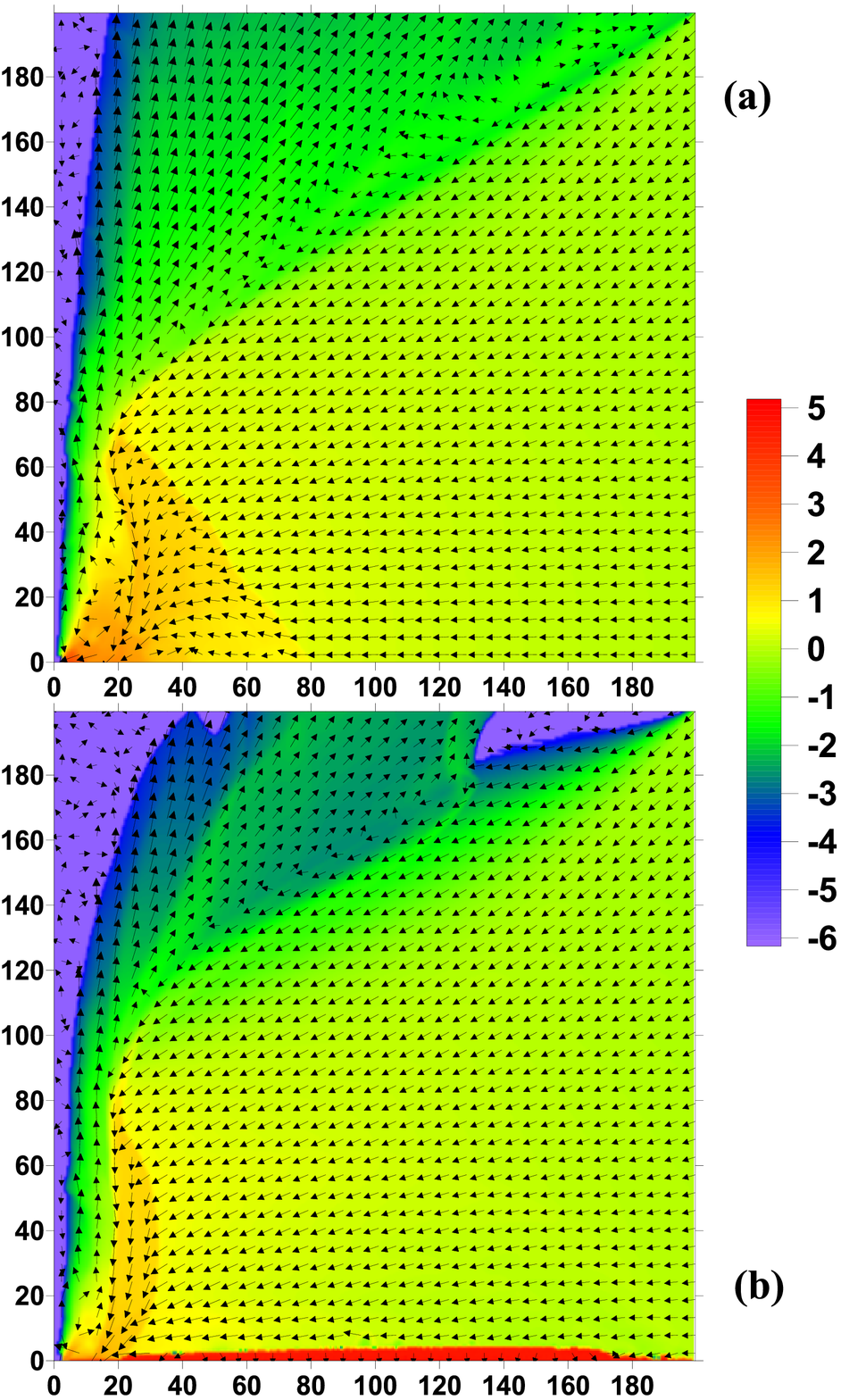,width=1.8in}
\psfig{file=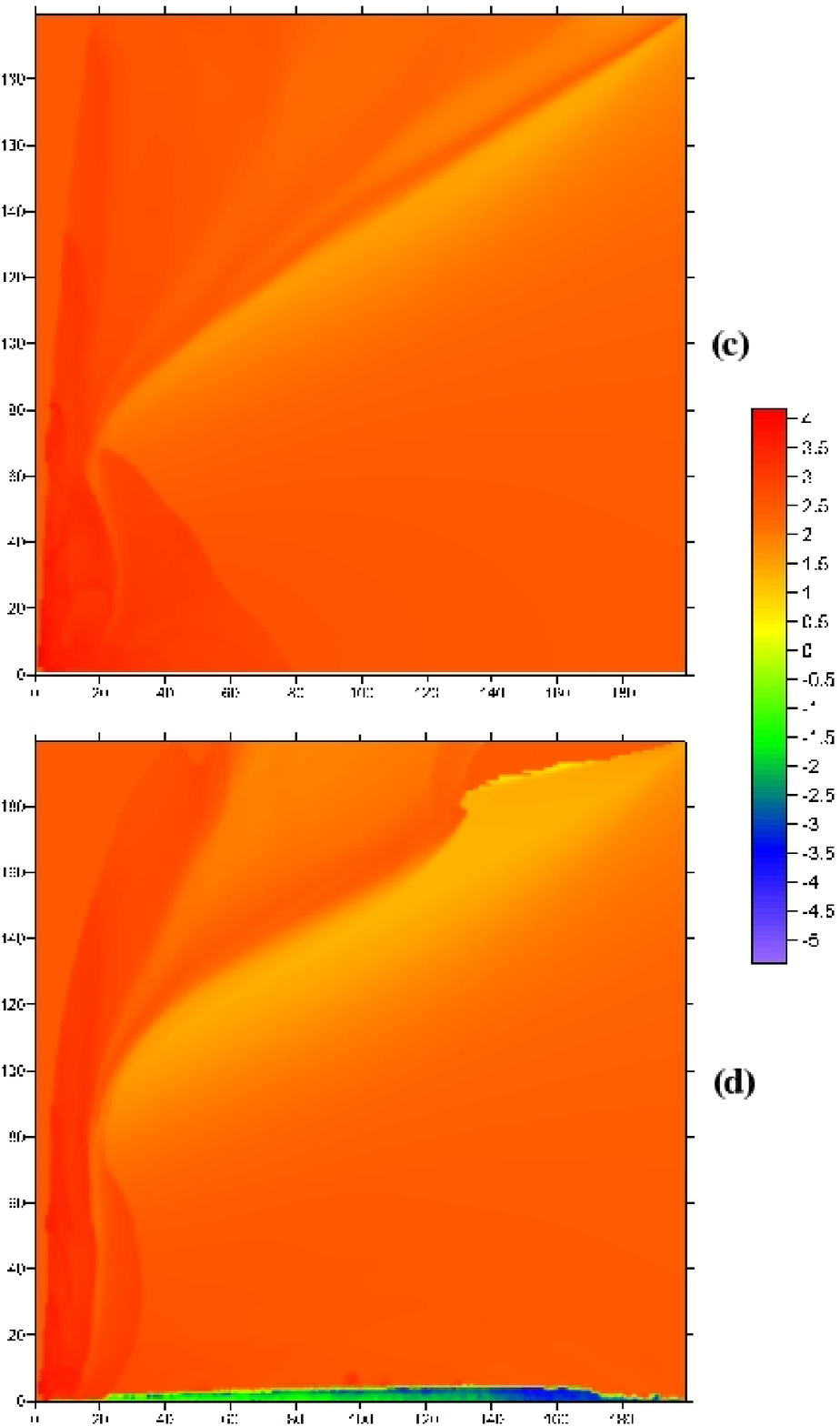,width=1.8in}
\end{center}
\caption{Changes in the $(i)$ density and velocity  $(ii)$ temperature distributions at $t= 95$s (a)/(c) without and (b)/(d) 
the inclusion of viscosity and cooling. Densities and temperatures in normalized units are plotted in logarithmic scale as in the scale on the right.
The density ranges from ${{log}_{10}}\rho = -6$ to $5$ in both the Figures.
A two component flow is clearly formed in (b) and (d).}
\end{figure}
Fig. 1(a-b) shows the velocity and the density distribution of the flow (a) without viscosity and cooling
and (b) with viscosity and cooling. In order to have a meaningful comparison,
all the runs were carried out up to $t=95$s. For both the cases, $\lambda = 1.7$
and ${\cal E} = 0.001$ were chosen. In Fig. 1b, we take ${\alpha}_{max} = 0.012$ 
The density distributions in Fig. 1(a-b) are plotted in the logarithmic scale shown on the right.
We note that on the equatorial plane, a Keplerian disk has formed out of the sub-Keplerian matter.
Close to the outer boundary, near the equatorial plane, we are injecting sub-Keplerian matter
and thus the Keplerian disk is disrupted there.
In Fig. 1(c-d), we show the temperature distributions
in keV as per color (logarithmic) scale on the right. In the absence
of cooling and viscosity, in Fig. 1c, the single component sub-Keplerian flow forms. In Fig. 1d,
because of higher viscosity, flows have the Keplerian distribution near the equatorial region.
Because of cooling effects, the region with a Keplerian distribution
is cooler and denser. Comparatively low dense sub-Keplerian  matter stays away from the equatorial plane.
For both the cases, CENBOL forms. 

\section{Concluding Remarks}
So far, there was no numerical simulations in the literature to
show that TCAF solution is realizable as a whole, and there was
no simulation to show whether such a configuration is at all stable.
Our result, for the first time, shows that if one assumes
that the viscosity is maximum on the equatorial plane, then, a low-
angular momentum injected flow is converted into a TCAF. We show
that the injected flow is segregated in the Keplerian and the sub-Keplerian components.
So far, we have captured all the salient features of the Keplerian
disk, by introducing a power-law cooling effect. In order to produce
an exact standard disk which emits multicolour black body as well,
we need to include the radiative transfer problem.
\section{Acknowledgments}
KG acknowledges financial support from both IUPAP and Prof. Remo Ruffini to attend MG13.

\end{document}